# How Does the Niagara Whirlpool Get Involved in Niagara River Self-Purification?

Anatoliy I Fisenko

*ONCFEC, Inc., 909 Lake Street, Suite 909, St. Catharines, Ontario L2R 5Z4, Canada*

*E-mail: afisenko@oncfec.com*

**Abstract.** The Niagara River self-purification through the natural formation of froth at a site of the Niagara Whirlpool basin has been investigated. It is shown that the naturally formed froth on the water surface of the Niagara Whirlpool contains a greater concentration of nutrients, trace metals and phenol in comparison to subsurface water. The natural pollutants removal process is explained in detail. As a result, the Niagara River at the Niagara Whirlpool basin possesses its own natural capacity to self-purify through the natural formation of froth. We can conclude that the Niagara Whirlpool is a natural source of Niagara River self-purification. The necessity for the long-term research on the sites of Niagara River for developing the remediation strategies is pointed out.

1. Introduction

It is well-known that rivers have a natural environmental capacity to self-purify their polluted sites, including the entire water and the benthic soil, through a complex of the chemical (oxidation, hydrolysis, photochemical reactions and others), the physical (sedimentation, evaporation, aeration and others), and the biological natural self-purification processes (Drinan and Spellman, 2001; Beyers and Odum, 1993; Grice and Reeve, 1982; Ostroumov, 1999; Klein, 1957; Vavilin, 1983; Loo and Rosenberg, 1989; Sommer, 1998; Hily, 1991; Mandi *et al.,* 1996; Robach *et al.,* 1991; Stimson *et al.* 1996; Otsuki *et al.* 1988; Logan and Hunt, 1987).

The biological self-purification processes play an essential role in the river self-purification via both the decomposition of total organic matter (natural and man-made) by fungi, bacteria and other microorganisms (Klein, 1957; Vavilin, 1983) and the utilizing of several functional biological filters (Loo and Rosenberg, 1989; Sommer, 1998; Hily, 1991; Mandi *et al.,* 1996; Robach *et al.,* 1991; Stimson *et al.* 1996; Otsuki *et al.* 1988; Logan and Hunt, 1987). In the process of decomposition, the entire water and the benthic soil are periodically enriched with a) biological surfactants such as amino, pyruvic, fatty and other acids; and b) the generated dissolved biogases – oxygen, ammonia, carbon dioxide and others.

Natural functions of the biological filters in the self-purification processes are the following: a) filtering water by filter-feeding organisms for removing phytoplankton and algae in the water, thus preventing the aquatic system from the rapid eutrophication (Loo and Rosenberg, 1989; Sommer, 1998; Hily, 1991). A typical example of these organisms are: rotifers, bryozoans, crustaceans, and others; b) filtering water by communities of aquatic plants in order to prevent the entry of pollution such as nitrogen and phosphorus from the surrounding land into stream water (Mandi *et al.,* 1996; Robach *et al.,* 1991); c) filtering water by benthic organisms for preventing the entry of polluting particles and biogenic elements into stream water from the benthic soil (Stimson *et al.* 1996). Some of the latter are *Tubifex, Chironomus, Asellus,* green macroalgae, *Dictyosphaeria cavernosa,* and others; d) filtering water by attaching microorganisms to the particles suspended in stream water (Otsuki *et al.* 1988; Logan and Hunt, 1987).

In earlier work (Fisenko, 2004), a new self-purification process of streams - natural froth formation was discovered and described. It was shown that the proper complex of the biological self-purification processes, such as the decomposition of total organics, together with the physico-chemical ones, caused by the proper level of turbulence, lead to the natural formation of the froth on a stream surface. The naturally formed froth collects all kinds of polluting particles, including organic, inorganic and the pathogenic bacteria, from the entire water body and the benthic soil. Based on the new insights into the stream self-purification, a new long-term in-stream on site clean-up method, dealing with the *non-point* (unregulated) sources of pollution has been proposed.

In subsequent works (Fisenko, 2006; Fisenko, 2008), the self-purification process through the natural froth formation has been studied for several streams. It was shown that the collected froth samples contain much greater concentration of the investigated polluting particles than water samples. These studies supported our idea that streams possess the self-purify activities through the natural formation of the froth.

The present paper is devoted to the subsequent development of the long-term in-stream on site remediation approach, which was previously proposed in (Fisenko, 2004). By way of example, a site of Niagara River at the Niagara Whirlpool basin (Ontario, Canada), where the natural formation of the froth takes place, has been studied. The preliminary test analysis showed that the naturally formed froth collects the tested polluting particles. As a result, the Niagara Whirlpool basin is a proper site where the natural self-purification of the Niagara River takes place.

2. Methods

On July 15, 2008, the subsurface water and the froth samples were collected downstream from Niagara Falls. The collecting site was a site at the Niagara Whirlpool basin. See Photo 1a.

**Photo 1:** Froth on the Niagara Whirlpool. 1a, on the water surface; 1b, on the bank.

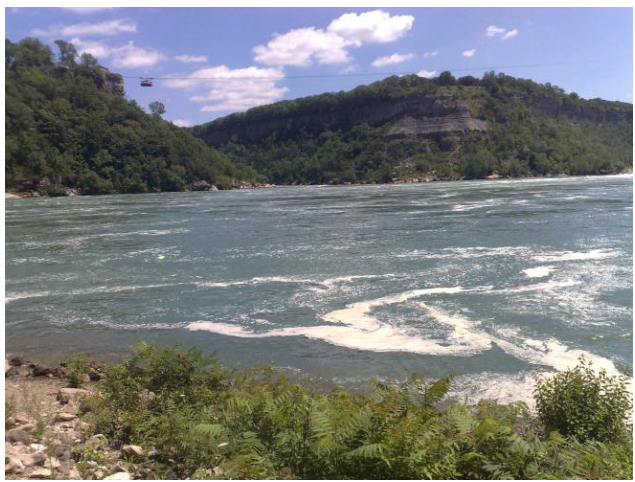 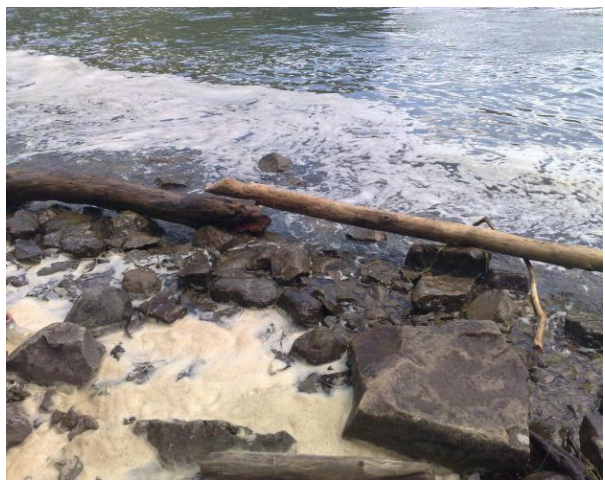

1a                                       1b

The Niagara Whirlpool is located about 6 km downstream from the Niagara Falls. The Niagara Whirlpool is a basin 518 m long by 365 m wide with depths up to 38 m (Niagara Falls and Great Gorge. Facts & Figures. (http://www.niagaraparks.com/)). The whirlpool is located in the middle of two sets of rapids where the Niagara River takes a ninety-degree turn and continues to flow towards Lake Ontario. The rapids upstream of the whirlpool have depths up to 15 m and the water speed can reach as high as 9 m/sec. The Niagara Whirlpool basin has the circular water motion clockwise. Below the whirlpool is another set of rapids, which drops about 12 m.

Subsurface water samples were collected in 0.5 L polyethylene bottles at about 15 cm depth. During sample collection the bottles were opened and closed at the sampling depth. Froth samples were collected from the bank of the whirlpool by using a polyethylene collector. See Photo 1b. Filtrations of the samples in which investigated polluting particles were measured were performed in the laboratory. For preparing samples for test analysis, the froth samples as well as the subsurface water samples were poured through filters with pores of 1 μm. One filter was necessary for filling of 10 mg of the filtered mixture from the froth samples. In this case, the particles larger than 1 μm had remained in the filters. Therefore, the test results were obtained for particles smaller than 1 μm in size. This range covers polluting particles in the colloid and the ionic states.

All samples were analyzed for nutrients, heavy metals, phenol, and turbidity by using the Palintest Photometer 5000 instrument (Palintest Ltd., www.palintest.com). The Palintest Photometer 5000 measures the color intensity, which is produced when reagents are added to the sample solution. The color intensity is proportional to the concentration of the investigated parameters under test. Reagents are supplied in the form of test tablets. The Photometer shows different chemical elements in the water samples in varying the color intensity. It is simple to use in field research and has been demonstrated to be very accurate.

### 3. Results and Discussion

Table 1 shows the comparison test results for concentrations of nutrients, heavy metals, phenol and turbidity in the froth and the subsurface water samples.

**Table 1** Comparison test results for samples taken from a site of the Niagara Whirlpool.

| CHEMICAL ELEMENTS: | Filtered Water From Subsurface Water Samples, mg/L | Filtered Mixture from Froth Samples, mg/L |
|---|---|---|
| **1) Turbidity**: | 5 FTU | 95 FTU |
| **2) Hexavalent Chromium ($Cr^{VI}$)** | 0.01 | 0.07 |
| **3) Aluminium (Al):** | 0.04 | 0.25 |
| **4) Nitrate (N):** | 0.21 | 0.58 |
| **5) Molybdate HR ($MoO4$):** | 1 | 4.2 |
| **6) Phenol ($C_6H_5OH$):** | 0.07 | 0.35 |
| **7) Phosphate ($PO_4$):** | 0.02 | 0.14 |

As clearly seen in Table 1, the concentrations of nutrients, trace metals and phenol in the filtered mixture, obtained from the froth samples, is much greater than those in the filtered subsurface water samples. Thus, phosphate rates seven times greater. Phenol rates five times greater. As for aluminium, its concentration is more than sixfold. As for turbidity, the concentration of undissolved substances in the froth samples rates nineteen times greater than in the subsurface water samples. Here it is important to note that the mixture from the froth contains contaminants in the suspended state and the heavily particles, too. This means that our test results do not provide full information regarding the concentration of pollutants in the froth. However, as clearly seen in Table 1, the concentration of the studied contaminants in colloid and ionic states is significantly higher than that in the subsurface water samples. As a result, the naturally formed froth collects the polluting particles from the Niagara River at the site of the Niagara Whirlpool.

Now let us consider the removal process of polluting particles by the natural formation of froth. It is well-known that for providing the flotation, the chemical reagents such as frothers and collectors should be added to water (Considine, D.M. and Cousidine, 1989). Frothers are surface-active chemicals, which form an adsorbing film on bubble surfaces, and the latter could obtain an electrical charge. Collectors are surface-active organic chemicals. They form an adsorbing film on particle surfaces and could, like the frothers, also obtain an electrical charge. In the Niagara River as well as in the Niagara Whirlpool basin, the flothers and the collectors are already presented. Indeed by decomposing the total organics in the entire water body and in/on the benthic soil by fungi, bacteria and other microorganisms, the biological surfactants such as amino, fatty and other acids are produced. Man-made surfactants such as detergents, soap and others in the river and the whirlpool are also present.

The proper level of turbulence or good mixing, created by the rapids upstream from the whirlpool, generates a large amount of dissolved air and air bubbles in river water. Then water enriched with the latter enters the whirlpool. Furthermore, the whirlpool basin itself is also enriched with dissolved biogases, generated during the decomposition processes therein. The generated air and biological bubbles could adsorb the available surface-active chemicals and obtain an electric charge. Besides, when the polluting particles adsorb the surface-active organic chemicals, they also obtain an electric charge. Upon being charged, the polluting particles are attached to the bubbles and form the bubble-particle aggregates. Then rise to the whirlpool water surface, the aggregates concentrate in the froth and the surrounding thin top layer of surface water. As a result, the resulting froth comprises a high concentration of the tested polluting agents such as nutrients, trace metals and phenol. By the circular water motion, the froth with a high concentration of polluting particles is moving around the whirlpool basin close to the bank and part of the froth concentrates on the bank. See Photo 1b. As the current moves the thinning froth mass around whirlpool, the bubbles in the froth gradually disappear. As a result, the heavier polluting particles could precipitate to the whirlpool benthic soil and the lighter ones with less amount of polluting particles is still in the whirlpool water body and flowing along the Niagara river downstream from the whirlpool.

As a result, the pollution removed from the Niagara River is concentrated in the Niagara Whirlpool basin. The latter allows a reduction in nutrients and a decrease in trace metals in the Niagara River. Therefore, the quality of river water that leaves the whirlpool is better than the water quality upstream from it. Consequently, the farther from the whirlpool, the lower will be the concentration of polluting particles in the river until a new kind of polluting sources (point or non-point) are encountered. This natural self-purification process through the froth formation should be work during a year. However, the verification of this hypothesis would require additional research.

The froth on rivers and creeks is caused by both natural processes and man-made pollution. The froth on river and creek surfaces results from a combination of the following: a) the presence of natural organic matter (e.g., dead fauna and flora), and/or industrial organic pollution; b) the activity of bacteria, fungi and other microorganisms decomposing the entire organic matter that is present in polluted sites of rivers and creeks: c) the inorganic pollution (e.g., detergents and chemical contaminations); and d) the turbulent water flow or good mixing. The amount of froth formed on a surface of the Niagara Whirlpool and on a surface of any river downstream from weirs, rapids, waterfalls and other obstacles, creating the shallow-turbulent character of water current strongly depends on the concentration of natural and man-made surfactants, the generated biogas and air bubbles as well as good mixing. The pollution removal effect in rivers will also depend upon these conditions.

In previous work (Fisenko, 2004) it was shown that the biological foam, created by fungi, bacteria and other microorganisms is a part of the natural froth. The biological foam plays an essential role in the stream self-purification. The similarity of processes involved in the foam formation in a stream and in activated sludge plants was pointed out in recent work (Fisenko, 2008). However, more research related to the identification of the proper groups of bacteria involved in the formation of froth in a stream should be conducted. This topic will be a point of discussion in subsequent publications.

In conclusion, it is important to note that there are several articles related to the testing of the froth and the subsurface water qualities on the Niagara River that support our results. As an example, the work done by the authors (Johnson *et al.,* 1989), the concentrations of the trace metals such as Cu, Cd, and Zn, the dissolved and the particulate carbon, and some classes of lipids were tested in the subsurface water and the froth samples taken below Niagara Falls. It was shown that the concentrations of the tested elements were much greater in the froth samples in comparison to the subsurface water ones. Similar results were obtained for a variety of radionuclides. (See (Platford and Joshi, 1989)).

## 4. Conclusions

The preliminary test analysis showed that the natural formed froth on a surface of the Niagara Whirlpool comprises a greater concentration of the investigated impurities in comparison to subsurface water. The latter collects pollutants, ranging from the nutrients, trace metals, phenol to the turbidity, taken from the Niagara River. The explanation of the contaminants removal process is considered in detail. Here we can conclude that the Niagara Whirlpool basin is the proper natural source of the Niagara River self-purification. At present, there is an essential interest in the expanding our research on the same site of the Niagara Whirlpool basin for obtaining a similar natural pollution removal effect throughout the year in order to observe this natural self-purification process by the froth formation at different seasons.

It is also very important to conduct a long-term research on other sites of Niagara River where the natural froth formation takes place. A special attention should be taken for the sites near Niagara Falls. In this case, research should be conducted for comparative analysis of the removed froth and water samples taken both the upstream and the downstream from Niagara Falls. As a result, a new long-term on site remediating approach for Niagara River could be developed. These and other topics will be points of discussion in subsequent publications.

## Acknowledgments

A special thanks to Mr. Brad Hill, the lead for Environment Canada's Niagara River Water Quality Monitoring Program, for his attention to our work done and the powerful discussions.